\documentclass[amsmath,amssymb,lengthcheck,aps,prl] {revtex4}
\usepackage{graphics}
\usepackage{graphicx}
\begin{document}

\title{ Roton minimum at $\nu=1/2$ filled fractional quantum Hall effect of Bose particles}

\author{Debashis Das}
\author{ Saswata Sahu }
\author{ Dwipesh Majumder }
\affiliation{Department of Physics, Indian Institute of Engineering Science and Technology, Howrah, WB, India}

\begin{abstract}
We have studied the collective excitation of fractional quantum Hall effect (FQHE) in the rotating Bose-Einstein condensate (BEC) using CF theory at the filling fraction $\nu=1/2$. The roton type of excitation in the FQHE of electron system is established over the years for all the filling fraction, whereas the collective excitation at $\nu=1/2$ filling fraction in the rotating BEC shows no roton minimum. We have investigated this using composite fermion theory with the P$\ddot{o}$schl-Teller interaction potential between the particles. We have seen that the long range interaction will give the roton, whereas short range interaction gives no roton minimum.

\end{abstract}
\maketitle 

%\section{Introduction}

Fractional quantum Hall effect (FQHE) \cite{fqhe} occurs due to the strong Coulomb interaction between electrons in two-dimensional system in presence of a perpendicular magnetic field. The natural quasiparticle  of FQHE is the composite fermion (CF)\cite{Jain_89}, each electron in the lowest Landau level (LLL) captures an even number of flux quanta and forms CF.  So the CF experience reduce amount of magnetic field, in this magnetic field CF forms new kind of Landau levels called $\Lambda$ levels, and most of the observed FQHE of strongly interacting electron can be mapped into the IQHE of non-interacting CFs. FQH states have reach collective phenomenon which has been studied over the past three and half decades.

If we rotate two dimensional harmonic trapped Bose-Einstein condensate (BEC)\cite{BEC}, above some critical angular velocity the motion of the superfluid forms vortex, if we increase the angular velocity we will have large number of vortices in triangular lattice arrangement\cite{vortexLatticTria, cooper2004}, which is also established by solving GP equation\cite{GPE} as we increase the rotation more and more vertices will generate and due to the increase of centrifugal force atoms will fly away of centre and the density of the fluid will reduce. In this low density, there is a theoretical proposal of the FQHE in rapidly rotating BEC of charge neutral dilute Bose gas in low temperature\cite{fqhe-bec,FQHE_BECref, FQHE_BECref1}. The neutral atoms do not interact with the magnetic field, but the rotation in the confinement potential plays a similar mathematical role of magnetic field in the two-dimensional electron system, sometimes this field refers as synthetic magnetic field\cite{Synthetic}. Atomic density of the system in the high rotation is very low, so we can assume that all the atoms will confine in the lowest LL.   %It is our theoretical interest to predict the nature of collective excitations in this system. 

%In case of fermions we have Pauli exclusion principle, but Boson do not follow any exclusion principle, so what will be the filling fraction ?

%We can think the FQHE in bosonic system in two different ways 1) 

% We have included all possible kind of interactions such as long range coulomb type interactions, dipole-dipole interactions and contact interactions.
 
  The CF realization of Bose atomic system is simply attaching odd number (say $p=1,3,5, \cdots$) of vortices with each atom\cite{CF_boson1, CF_boson2, rot_less4}. The magnetic field experience by the CF of Bose particle is
\begin{eqnarray}
B^* = B - p \rho \phi_0
\end{eqnarray}
where $B$ is the actual fictitious magnetic field, $\rho$ is the number density of the Bose particle, $\phi_0$ is the magnetic flux quantum. 
In this reduced  magnetic field CF forms Landau level, called $\Lambda$ levels. 
The lowest LL filling fraction of boson and filling fraction of CF ($n$, an integer number of filled $\Lambda$ level) is given by
\begin{eqnarray}
\nu = \frac{n}{np + 1}
\end{eqnarray}

In FQHE of electron, $\nu$ is the ratio of the number density of electrons to the degeneracy per unit area of LL, here in rotating trap $\nu$ is the ratio of the number of bosons to the average number of vortices.
The ground state properties of the Bosonic counterpart of FQHE have been studied but little attention has been given to the collective excitation in this system in the thermodynamic limit. There are some exact diagonalization calculation and CF calculation for small number of particles.
 The roton type of excitation in the FQHE of electronic system is a natural phenomenon, which is established theoretically as well as experimentally\cite{roton-electron1,roton-electron2}, roton minimum has been predicted theoretically in the collective excitation of filled LL of dipolar interacting fermions\cite{roton_dipole}. It has been observed that the excitation spectrum FQHE of Bosonic system at filling fraction $\nu=1/2$ is not associated with the roton type of excitation\cite{CF_boson1,rot_less2,rot_less3,rot_less4}, whereas some other filling fractions contain roton-minimum in their energy spectrum. In this article, we have tried to explain the absence of roton minimum in the energy spectrum at the filling fraction $\nu=1/2$. Instated of considering delta function potential we have considered short-range Poschl-Teller(PT) interaction between atoms. In addition of PT interaction, we have considered coulomb interaction too\cite{chargedBoson} to see the nature of excitations in the long-range interaction.

\section*{P$\ddot{o}$schl-Teller  Interaction between dilute Bose gas atoms}
In the most of the cases the effective interaction between
two charge neutral Bose particles at low energies is a constant in the momentum representation, $U_ 0 = 4\pi  \hbar^2 a/m$, where $m$ is the mass of each particle and $a$ is s-wave scattering length. In the real space the interaction can be expressed as
\begin{eqnarray}
V = U_0 \sum_{i<j} \delta^{(3)}(\vec{r}_i - \vec{r}_j)
\end{eqnarray}
This interaction gives the well known GP equation for BEC and superfluid system. In two dimension this expression become
\begin{eqnarray}
V = g \sum_{i<j} \delta^{(2)}(\vec{r}_i - \vec{r}_j)
\end{eqnarray}

The interaction strength can be tuned by changing scattering length $a$ in presence of magnetic field. It is very difficult to handle the delta function potential in quantum Monte Carlo calculation and it also requires huge computational resource. Using delta function potential it is not possible to calculate the energy spectra for large number of particles\cite{rot_less4}. To avoid this difficulty and to access the system size in the thermodynamic limit we have considered P$\ddot{o}$schl-Teller interaction (PT) potential\cite{Poschl-Teller}
\begin{eqnarray}
V = g \sum_{i<j} \frac{2\mu}{cosh^2 \;( \mu r_{ij})}
\end{eqnarray}

\begin{figure}
  \includegraphics[width=11cm]{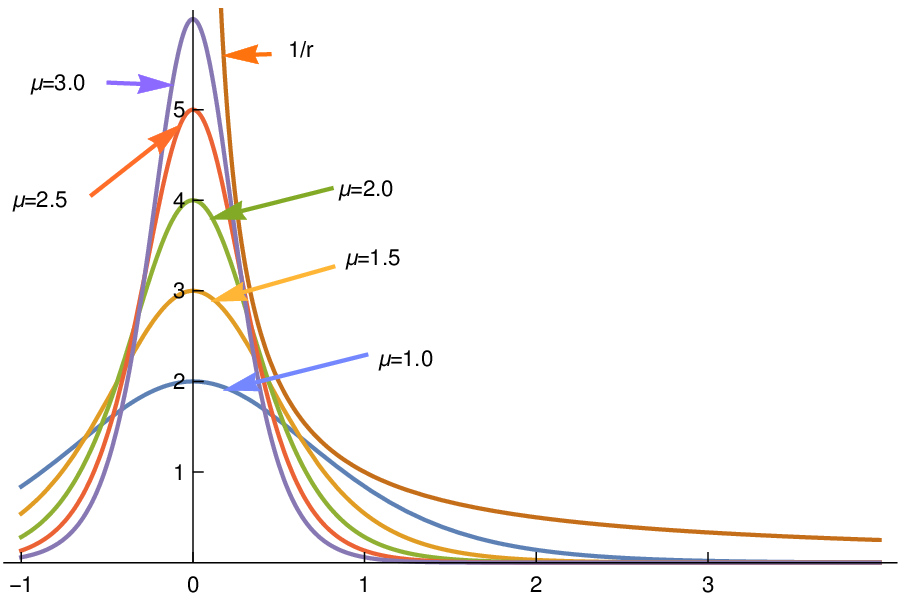}
  \caption{P$\ddot{o}$schl-Teller interaction potential, $\frac{2\mu}{cosh^2 \;( \mu r)}$ as function of separation distance of two particles in unit of magnetic length ($l$) for different values of $\mu$ in unit of inverse of $l$. We ploted the potential for both side to see the delta function nature of the interaction, though the value of $r$ is sholly positive.  As we increase the value of $\mu$ the potential become more and more like delta-function potential. The out most line is the 1/r plot for comparison.}
  \label{deltaPotl}
\end{figure}

where 1/$\mu$ is the width of the interaction, $\mu$ is the parameter of interaction in unit of inverse of the magnetic length $l=\sqrt{\hbar c/eB} = \sqrt{\hbar/ m \omega}$. Here we have considered a range of $\mu$, to investigate the $\mu$ dependence nature of excitation. Small value of $\mu$ gives flat nature of potential as we increase the value of $\mu$ the nature of potential become delta type. Very large value of $\mu$ will give zero energy as the average separation between particles will be large compared to the width of interaction. This PT interaction gives us the opportunity to study the FQHE from long range to short range of interaction.

\section*{Wave function \& calculation procedures}

The standard spherical geometry is used in our calculations, which
considers electrons moving on the surface of a sphere, subjected
to a radial magnetic field. The magnetic field can be thought
to emanate from a ‘magnetic monopole’ of strength Q at the
centre, which produces a total magnetic flux of $2Q\phi_0$ through
the surface of the sphere of radius $R = \sqrt{Q} l$. This maps into a system of composite
fermions at an effective flux $ 2q = 2Q - (N - 1)$, with $Q$ chosen
so that the state at $q$ is an integral quantum Hall state at
filling $n = 1$ so that we will have $\nu = 1/2$ filling fraction. In spherical geometry, the angular momentum number is a good quantum number and its value of an electron in the k-th LL is $k+Q-1$\cite{monopole}.

The ground state wave function of the $N$ electron system of FQHE at filling fraction $\nu$, which maps with $n$ filled $\Lambda$ levels of CFs is\cite{Jain_89,Wu_Jain2013}
\begin{equation}
  \Psi^0=J^{-1} P_{LLL} J^2 \; \Phi_1(\Omega_1, \Omega_2, \cdots \Omega_N)
\end{equation}
where $\Omega_i$ are the position of electron on the surface of the sphere, $\Phi_1$ is the Slater determinant of complitely filled lowest $\Lambda$ level of CFs, $P_{LLL}$ is the projection operator onto the LLL and the Jastrow factor is given by
\begin{eqnarray}
  J = \prod _{i<j}^N (u_i v_j - u_j v_i)
\end{eqnarray}
where the spinor variables are $u=cos(\theta/2) \; exp(-i\phi /2)$ and $v=sin(\theta/2) \; exp(i\phi /2)$ with $0\le \theta \le \pi$ and $0\le \phi \le  2\pi$. Here $\Phi_1$ and $J$ both are odd under the exchange of particles so the wave function of the system remains symmetric after the fermionic transformation. 
The ground state wave function of $\nu=1/2$ is already in the LLL, but the excited states need the projection.

% only for 1\2 filling 
The excited state wave function of $N$-particle system at the filling fraction $\nu=1/2$ corresponding to the transition of a CF from the filled 0th $\Lambda$ level to an empty $\Lambda$ level $n_f\; (n_f> 0)$, in the spherical geometry is given by \cite{Kamilla97, DM14, Scarola}
{\small{
\begin{equation}
  \Psi (L)=J^{-1} P_{LLL} J^2 \sum_{m_h} |m_h>\; <q, m_h; n_f+q,m_p|L,M>
\end{equation}}}
where $|m_h>$ is the Slater determinant of $N-1$ number of particles in the lowest $\Lambda$ level with a hole at $m_h$  Z-component of angular and one particle in the $n_f$ $\Lambda$ level with Z-component of angular momentum $m_p$, $<q, m_h;n_f+q,m_p|L,M>$ are the Clebsch-Gordan coefficients, $L$ is the total angular momentum

  Here we have considered sub-Hilbert space with zero Z-component of angular momentum ($M=0$) without any loss of generality to reduce the numerical complication.
 Actual collective excitation is not a single CF-exciton state rather the superposition of all possible excitons. We have presented the results of three excitons as we have checked that this energy is identical with the calculation considering four excitons. The excitons are not orthogonal, we have used Gram-Schmidt Orthonormalization procedure to orthogonalize low energy exciton states with a fixed angular momentum. The method of calculation of energy of such kind of mixed state is called CF-diagonalization\cite{CFD}.

The excited state energy with respect to the  ground state  $\Psi^0$  is given by 
\begin{equation}
  \Delta(L) = \frac{<\Psi(L)| H |\Psi(L)>}{<\Psi(L) |\Psi(L)>} - \frac{<\Psi^0| H |\Psi^0 >}{<\Psi^0 | \Psi^0 >}
\end{equation}
$H$ is Hamiltonian of the system. As the kinetic energy become quantized and we assume that the particles are confined in the LLL, the Hamiltonian of the system will be $H=V$. The multidimensional integration has been carried out using quantum Monte Carlo method.

\section*{Results \& discussion}

\begin{figure}
 \includegraphics[width=8cm]{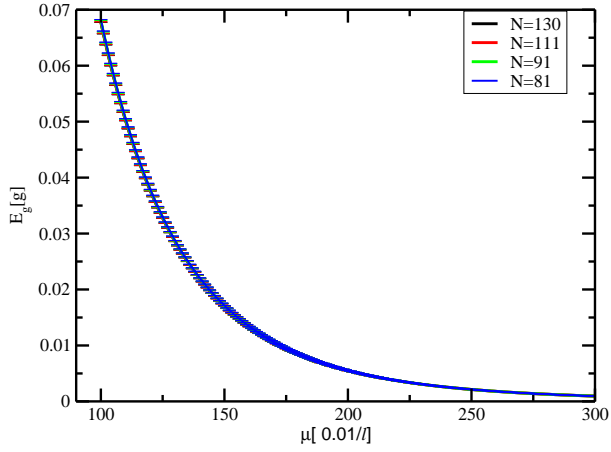}
 \caption{ Ground state energy per particle in unit of g as a function of $\mu$, for different particle number.} 
 \label{ground}
\end{figure}

We have calculated the ground state energy per particles in a range of interaction parameter $\mu$ from 1.0 to 3.0 for different number of particles $N=$130, 111, 100, 91 and 81 as shown in the FIG \ref{ground}. The result shows that the number of particles is sufficient to get the thermodynamic limit of physical quantities. 

The energy spectrum for different values of $\mu$ has been shown in the FIG \ref{spec_del}. We have used the same number of particles to calculate the excited state energy, the average energy has been shown in the figure. Small value of $\mu$ gives sharp roton minimum, as we decrease the range of interaction by increasing the interaction parameter $\mu$ the roton minimum becomes shallow and vanishes at some limit of interaction.

\begin{figure}
  \includegraphics[width=8cm]{rot_st1.eps}
  \caption{Energy spectra for different values of $\mu$.The arrow line represents the ascending order of $\mu$. At low value of $\mu$, ie at long range interaction we have a very sharp roton minimum, as we increase the value of $\mu$ ie decrease the range of interaction the roton minimum becomes shallow and vanishes after some limiting range of interaction. We have calculated the energy spectra for 131, 111 and 91 number of particles. Our result is accurate upto fourth decimal points.}
  \label{spec_del}
\end{figure}

\begin{figure}
  \includegraphics[width=8cm]{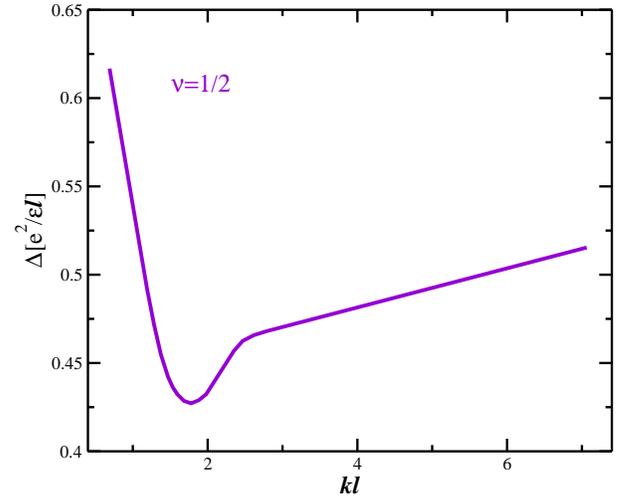}
  \caption{Energy spectra for coulomb interaction}%Unit of energy is $e^2/\epsilon l$, where $l$ is the magnetic length.Wave vector is related to the total angular momentum by $kl = L/R$  .}
  \label{coulomb}
\end{figure}
%In figure 4 we have ploted the energy spectrum for the coulomb interaction at filling fraction $\nu=1/2$. There is a sharp minimum in the energy spectra, roton minimum.

So the nature of the roton minimum is due to the short range of interaction at the filling fraction $\nu=1/2$. Our energy is slightly higher than the energy spectra of delta function interaction. We could not able to go as close as delta function interaction as the ground state energy becomes very small. We see that as we are going to the short range region the energy is reducing and we believe that the energy will be identical with the delta function interaction energy spectra. 

The interaction responsible for the FQHE in electronic system is the Coulomb interaction, which is long range in nature. We have calculated the energy spectra for Coulomb interaction FIG \ref{coulomb} and we see that there is a sharp roton minimum in the energy spectra. So we confirm that the disappearance of the roton minimum is due to the short range of interaction.

\section*{Acknowledgement}

Debashis thanks UGC, India for the financial support.


\begin{thebibliography}{9}

 \bibitem{fqhe} D. C. Tsui, H. L. Stormer and A. C. Gossard, Phys. Rev. Lett. {\bf 48}, 1559 (1982), Phys. Rev. B {\bf 25}, 1405 (1982) ; H. L. Stormer, A. Chang, D. C. Tsui, J. C. M. Hwang, A. C. Gossard and W. Wiegmann, Phys. Rev. Lett. {\bf 50}, 1953 (1983).
\bibitem{Jain_89}J. K. Jain, Phys. Rev. Lett. \textbf{63}, 199 (1989), Phys. Rev. B \textbf{41}, 7653 (1990).


\bibitem{BEC}     M. H. Anderson, J. R. Ensher, M. R. Matthews, C. E. Wieman, E. A. Cornell, Science Vol. 269, Issue 5221, pp. 198 (1995);   K. B. Davis \textit{et al.}, Phys. Rev. Lett. \textbf {75}, 3969(1995); 

\bibitem{vortexLatticTria} K. W. Madison, F. Chevy, W. Wohlleben and J. Dalibard, Phys. Rev. Lett. {\bf 84}, 806(2000); M. R. Matthews, B. P. Anderson, P. C. Haljan, D. S. Hall, C. E. Wieman and E. A. Cornell, Phys. Rev. Lett. {\bf 83} , 2498(1999); J. R. Ensher, D. S. Jin, M. R. Matthews, C. E. Wieman and E. A. Cornell, Phys. Rev. Lett. {\bf 77} , 4984(1996).

\bibitem{cooper2004} N. R. Cooper, S. Komineas, N. Reed, HYSICAL REVIEW A 70, 033604 (2004); N. R. Cooper, N. K. Wilkin and J. M. F. Gunn hysRevLett.87.120405(2001). 

\bibitem{GPE} Bao W. (2007) Ground states and dynamics of rotating BECs. In: Cercignani C., Gabetta E. (eds); Weizhu Bao, Hanquan Wang, J. Comput. Phys. \textbf{217}, 612(2006)


\bibitem{fqhe-bec} N. K. Wilkin, J. M. F. Gunn, and R. A. Smith, Phys. Rev. Lett. {\bf 80}, 2265 (1998)
\bibitem{FQHE_BECref} N. Regnault and Th. Jolicoeur Phys. Rev. Lett.{\bf 91} , 030402 (2003)  \&  Phys. Rev. B {\bf 69}, 235309 (2004); Susanne Viefers, J. Phys. - Cond. Mat.{\bf 20} , 123202 (2008); S. Viefers, T. H. Hansson, and S. M. Reimann, Phys. Rev. A {\bf 62}, 053604 (2000); 
\bibitem{FQHE_BECref1} S. Furukawa and M. Ueda Phys. Rev. A {\bf 96}, 053626(2017)
\bibitem{Synthetic} Y. Lin, Rob L. Compton, K. J. Garcia, J. V. Porto, I. B. Spielman, Nature {\bf 462}, 628 (2009).

\bibitem{CF_boson1} N. R. Cooper and N. K. Wilkin, Phys. Rev. B {\bf 60}, R16279 (1999).
\bibitem{CF_boson2} N. Regnault, C. C. Chang, T. Jolicoeur and J. K. Jain,  J. Phys. B {\bf 39}, S89 (2006)\& J. Phys. B: At. Mol. Opt. Phys. 39 (2006) S89--S99. 

\bibitem{rot_less4} C. C. Chang, N. Regnault, T. Jolicoeur, and J. K. Jain, Phys. Rev. A {\bf 72}, 013611 (2005)


\bibitem{roton-electron1} J. Tersoff, Phys. Rev. B {\bf 30} , 4655(1984);  J. E. Demuth and B. N. J. Persson, Phys. Rev. Lett. {\bf 54} , 584(1985);  G. Dev and J. K. Jain, Phys. Rev. Lett.   {\bf 69}, 2843 (1992);  V. W. Scarola, K. Park, and J. K. Jain, Phys. Rev. B {\bf 61} , 13064(2000); K. Park, J.K. Jain, Solid state comm. {\bf 115}, 353(2000);

\bibitem{roton-electron2}  M. A. Eriksson, A. Pinczuk, B. S. Dennis, S. H. Simon, L. N. Pfeiffer, and K. W. West, Phys. Rev. Lett. {\bf 82}, 2163 (1999);     I. V. Kukushkin, J. H. Smet, V. W. Scarola, V. Umansky, K. v. Klitzing, Science 324, Issue 5930, 1044(2009).

\bibitem{roton_dipole} J. N. Wu, Y. L. Chiang, H. C. Huang and S. C. Cheng, J. Phys. B: At Mol. Opt. Phys. 44, 245301(2011); Shih-Da Jheng, T. F. Jiang and Szu-Cheng Cheng, PHYSICAL REVIEW A 88, 051601(R) (2013).

%\bibitem{rot_less1}N. R. Cooper and N. K. Wilkin,Phys. Rev. B {\bf 60} , R16279(1999)

\bibitem{rot_less2} N. Regnault and T. Jolicoeur, Phys. Rev. Lett. {\bf 91} , 030402 (2003)

\bibitem{rot_less3} N. Regnault and T. Jolicoeur,Phys. Rev. B  {\bf 69} , 235309 (2004)



\bibitem{chargedBoson} S. K. Ma and C. W. Woo, Theory of a Charged Bose Gas. I, Phys. Rev. {\bf 159}, 165 (1967); C. W. Woo and S. K. Ma, Theory of a Charged Bose Gas. II, Phys. Rev. {\bf 159}, 176 (1967)

\bibitem{Poschl-Teller} A J Morris, P L Rios, R J Needs, Phys. Rev. A {\bf 81}, 033619 (2010).

\bibitem{monopole} T.T. Wu and C.N. Yang, Nucl. Phys. B {\bf 107}, 365 (1976); T.T. Wu and C.N. Yang, Phys.
Rev. D{\bf 16}, 1018 (1977).


\bibitem{Wu_Jain2013} Y. H. Wu and J. K. Jain, Phys. Rev. B {\bf 87}, 245123 (2013). 


\bibitem{Kamilla97} J. K. Jain and R. K. Kamilla, Int. J. Mod. Phys. B {\bf 11}, 2621 (1997).
\bibitem{DM14} D. Majumder, S. S. Mandal and J. K. Jain, Nat. Phys. {\bf 5}, 403 (2009); D. Majumder and S. S. Mandal Phys. Rev. B {\bf 90}, 155310 (2014).
\bibitem{Scarola} V. W. Scarola, K. Park and J. K. Jain, Phys. Rev. B {\bf 61}, 13064 (2000).




\bibitem{CFD}S. S. Mandal, J. K. Jain, Phys. Rev. B {\bf 66} , 155302 (2002).

%\bibitem{review1} Quantum Hall physics in rotating Bose–Einstein condensates, Susanne Viefers, Journal of Physics: Condensed Matter {\bf 20}, 12 (2008)


\bibitem{2include Long-Short-range} Rapidly rotating boson molecules with long- or short-range repulsion: An exact diagonalization study, Leslie O. Baksmaty, Constantine Yannouleas, and Uzi Landman, Phys. Rev. A 75, 023620 – Published 26 February 2007







%\bibitem{dipole1} A. Griesmaier, J. Werner, S. Hensler, J. Stuhler, and T. Pfau, Phys. Rev. Lett. {\bf 94}, 160401(2005)
%\bibitem{dipole1A} J. Stuhler, A. Griesmaier, T. Koch, M. Fattori, T. Pfau, S. Giovanazzi, P. Pedri, and L. Santos, Observation of Dipole-Dipole Interaction in a Degenerate Quantum Gas, Phys. Rev. Lett. {\bf 95}, 150406 (2005).
%\bibitem{dipole2} T. Lahaye, T. Koch, B. Frölich, M. Fattori, J. Metz, A. Gries- maier, S. Giovanazzi, and T. Pfau, Nature ͑London͒ {\bf 448}, 672 (2007)
%\bibitem{dipole3}Fermions out of dipolar bosons in the lowest Landau level B. Chung and Th. Jolicoeur, PRA {\bf 77}, 043608 (2008)
%\bibitem{dipole4} Abraham J. Olson, * Daniel L. Whitenack, and Yong P. Chen, PHYSICAL REVIEW A {\bf 88} , 043609 (2013) Effects of magnetic dipole-dipole interactions in atomic Bose-Einstein condensates with tunable s-wave interactions
%\bibitem{dipole5}Kui-Tian Xi 1 and Hiroki Saito, PRA{\bf 93} , 011604(R) (2016) Droplet formation in a Bose-Einstein condensate with strong dipole-dipole interaction




%K. W. Madison, F. Chevy, W. Wohlleben and J. Dalibard, Phys. Rev. Lett. \textbf{84}, 806(2000).


%\bibitem{BoseStar} G. Ingrosso and R. Ruffini, On Systems of Self-Gravitating Bosons and Fermions Undergoing Quantum Condensation. Newtonian Approach, Il Nuovo Cimento 101B, 369 (1988)






\end{thebibliography}
\end{document}